\documentclass[conference]{IEEEtran}
\IEEEoverridecommandlockouts
\usepackage{cite}
\usepackage{amsmath,amssymb,amsfonts}
\usepackage{algorithmic}
\usepackage{graphicx}
\usepackage{textcomp}
\usepackage{xcolor}
\usepackage{tabularx}
\usepackage{algorithm2e}
\usepackage{lipsum}  
\usepackage{float}

\RestyleAlgo{ruled}
\def\BibTeX{{\rm B\kern-.05em{\sc i\kern-.025em b}\kern-.08em
    T\kern-.1667em\lower.7ex\hbox{E}\kern-.125emX}}
\begin{document}

\title{Active Connectivity Fundamentals for TSCH Networks of Mobile Robots\\}

\author{\IEEEauthorblockN{Charalampos Orfanidis, Paul Pop, Xenofon Fafoutis}
\IEEEauthorblockA{Department of Applied Mathematics and Computer Science, Technical University of Denmark \\
{\{\fontfamily{qcr}\selectfont chaorf,paupo,xefa\}}@dtu.dk}

}

\maketitle

\begin{abstract}
Time Slotted Channel Hopping (TSCH) is a medium access protocol defined in the IEEE 802.15.4 standard which have been proven to be one of the most reliable options when it comes to industrial applications. TSCH has been designed to be utilized in static network topologies. Thus, if an application scenario requires a mobile network topology, TSCH does not perform reliably. In this paper we introduce active connectivity for mobile application scenarios, such as mobile robots. This is a feature that enables the option to regulate physical characteristics such as the speed of a node as it moves, in order to keep being connected to the TSCH network. We model the active connectivity approach through a basic example where two nodes are moving towards the same direction to infer the main principles of the introduced approach. We evaluate the active connectivity feature through simulations and quantify trade-off between connectivity and application-layer performance.

\end{abstract}

\begin{IEEEkeywords}
TSCH, Mobility, Dependable IoT, Reliability.
\end{IEEEkeywords}

\section{Introduction}
\label{sec:introduction}

Time Slotted Channel hopping (TSCH) \cite{Duquennoy2017} is a Medium Access Control (MAC) protocol which became very popular in Industrial Internet of Things (IoT) \cite{8401919} since it can provide robustness and reliability towards the network performance. The TSCH protocol is based on a channel hopping function that mitigates the impact of interference and multipath fading that diminish the communication in IoT networks. All the nodes in TSCH networks share a common time source to enable synchronization among the nodes. Provided that the nodes are synchronized, a schedule is calculated including a predefined channel offset and time offset which declares the communication details within the TSCH network. 

TSCH is a flexible protocol which can be tailored towards the requirements of each application. A very popular customization approach is modifying the TSCH schedule. Since the traffic demands, network topology and routing might vary a lot depending on the application, the TSCH schedule is a crucial part which can be adjusted based on the application demands. Therefore, there is an extended scientific literature around TSCH scheduling schemes, which includes three main categories: centralized \cite{WirelessHART}, distributed \cite{rfc4180} and autonomous \cite{Duquennoy2015}. 

As TSCH networks emerged, they started being utilized in different kind of dependable applications with mobile nodes, that each one has various set of requirements with different challenges every time. For example, wearable sensors \cite{10.1145/3366617} for health, are mobile devices which could benefit from a reliable mobile TSCH. 
In \cite{6lowpan}, the node mobility is distinguished between macro and micro mobility: macro is the mobility between different network domains and micro is the mobility within the current network domain. In this paper we focus on micro mobility and more specifically mobility within the TSCH network. In our previous work \cite{iot2040033} we demonstrated that the current state of the art in TSCH schedules cannot support mobility. Thus, TSCH performance is far from reliable when mobile communication is required. There are several factors to be taken into account in a mobile scenario like the traffic volume, the mobility model, the power budget and the operating environment. 

Mobile robotics is another application that could benefit from reliable TSCH-based robot-to-robot communication, such as Unmanned Aerial Vehicles (UAV) \cite{8993742}, 
autonomous vehicles \cite{8501581} and many more. 
Indeed, similar to wireless sensor networks, mobile robots are cyber-physical systems that depend on traditional (\textit{e.g.}, \cite{jayasuriya2020active}) and nontraditional sensors (\textit{e.g.}, \cite{marchegiani2018learning}) to perceive the environment and act to maximise their application objectives. This synergy is recently investigated as part of the Internet of Robotic Things \cite{simoens2018internet,kamilaris2020penetration}.

Unlike wearable sensors, whereby the device has no control on the mobility of the user,  mobile robots routinely adjust their mobility to maximize their objectives, e.g. planning based on energy consumption estimates \cite{7759348,7139866}. When reliable communication is vital, a mobile robot can make similar adjustments for maintaining reliable connectivity based on link quality estimates.
In this spirit, we introduce the concept of Active Connectivity (AC). Namely, a node has the ability to regulate its physical characteristics (\textit{i.e.}, speed, route, topology) to remain within coverage of its neighbours and therefore keep being connected to the TSCH network. While this approach can increase the reliability and robustness of the TSCH network, it may have negative impact on other objectives, such as the time the robot needs to cover a certain distance in order to complete a task and consequently to the overall cost. 
Thus, we investigate the tradeoff between the reliable connectivity and application performance in terms of distance covered. In order to acquire a fundamental insight into the performance of AC we opted for studying a basic scenario, whereby two mobile nodes move towards the same direction. For the scope of the evaluation we used the Cooja simulator \cite{cooja}, a cross layer network simulator which includes different levels from physical to application layer.

This paper is organized as follows. A brief technical description of TSCH is mentioned in Section \ref{sec:tsc_primer}. Section~\ref{sec:Relate_work} presents the state of the art and how it is related with the current work. Section \ref{sec:active_connectivity} illustrates the AC operating principles. The experimental scenario, the evaluation details the obtained results and a discussion are presented in Section \ref{sec:evaluation}. Finally Section \ref{sec:Conclusion} concludes the paper and discusses future work.

\section{TSCH Primer}
\label{sec:tsc_primer}
This Section provides a brief overview with the technical aspects of TSCH protocol which is part of the IEEE 802.15.4 standard \cite{IEEE802.15.4} and an amendment in IEEE 802.15.4e-2015 \cite{IEEE802.15.4e}.

\begin{figure}[tp]
    \centering
    \includegraphics[scale=.7]{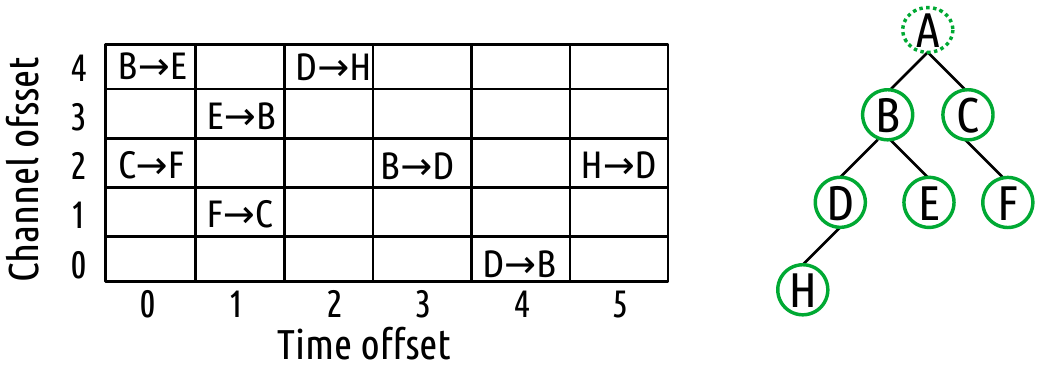}
    \caption{TSCH slotframe on the left and routing topology on the right.}
    \label{fig:TSCH_background}
\end{figure}

TSCH operates at the link layer and two of its main features is the frequency hopping and the synchronization of the nodes. The nodes of the TSCH network are following a duty cycle which is dictated by the schedule. They have the option to sleep for reserving power, transmit or receive a packet. In order to avoid collisions, it is not allowed to multiple transmissions at the same time and channel. The schedule is organized a two-dimensional table called the slotframe. Figure $\ref{fig:TSCH_background}$ illustrates a slotframe with a topology. The time offset is represented on the x-axis and the channel offset at the y-axis. Every cell, which is called timeslot, represents a specific time and channel offset and most of the times has a duration of $10$ ms which is enough to send an IEEE 802.15.4 packet and receive an ACK. In Figure $\ref{fig:TSCH_background}$ there are $6$ timeslots and $5$ channel offsets. The number of the timeslots is defining the length of the slotframe and has an impact on several tradeoffs affecting the network performance. 

The TSCH coordinator is constructing the TSCH network, it defines the network ID, slotframe size and the Frequency Hopping Sequence (FHS). The coordinator is also responsible to initialize the Absolute Slot Number (ASN) to $0$ which is increased with a new timeslot. To join the TSCH network a node has to receive an Enhanced Beacon (EB) packet, which is broadcast from the already joined nodes. EB contains essential information like the time source which is necessary to synchronize with the other nodes. Equation $\ref{eq:channel}$ is computing a channel for a given cell, CO stands for Channel Offset. There is a priority mechanism in case there is more than one transmission cell at the same time.

\begin{equation}
    Channel = FHS(ASN + CO)\; mod\; ||FHS||
\label{eq:channel}
\end{equation}

\subsection{Minimal Scheduling Function (MSF)}
During the evaluation part we used the MSF schedule to keep this part simple as well. The MSF is defined in the RFC 9033 \cite{rfc4180} as a scheduling mechanism for TSCH implemented at the top of 6top protocol. MSF include three type of cells: minimal, autonomous, negotiated. Minimal cells are used for EB packets and routing purposes; autonomous are used for unicast communication and a negotiated cell is used when a node has just joined the TSCH network and intends to let the other nodes about its presence. 

\section{Related Work}
\label{sec:Relate_work}

Utilizing TSCH under mobile scenarios is a research topic which is not covered adequately by the research literature. This section illustrates the existing research endeavours but also a few instances presenting how RPL (Routing Protocol for Low-Power and Lossy Networks) performs in mobile scenarios as well. RPL is used in several TSCH approaches as the default routing protocol and information obtained from RPL are used to construct the TSCH schedule at some case. Therefore, it is interesting to understand how it performs on mobile scenarios and what may be its impact on TSCH.

Al-Nidawi et al \cite{AlNidawi2015}, evaluated the Low Latency Deterministic Network (LLDN) and TSCH protocols under mobility scenarios. The authors identify four main issues TSCH that affect network performance and quantify their impact on the energy consumption and the network coverage. The main issues mentioned are the long channel offset which increases the scanning process, mobile nodes cannot identify coordinators due to beaconing in multiple frequencies and the absence of a flexible timeslot scheme which will enable a smoother function when a node is added or deleted. After gaining an insight into the challenges of TSCH in mobile scenarios Al-Nidawi et al.     \cite{Nidawi_b} propose Mobile TSCH to decrease the delay that arises during the TSCH joining procedure. In order to achieve this they encapsulate Enhanced Beacon (EB) packets into ACK packets which are transmitted on a fixed channel. The proposed approach decreases the Radio Duty Cycle (RDC) by an average of $30$\% and increases the connectivity by $25$\%. An evaluation of TSCH in mobile scenarios which considers both mobile and stationary nodes presents that TSCH can can achieve adequate connectivity with the premise that there are adequate amount of nodes either stationary or mobile to enable a proper coverage \cite{Raza2019}. It is introduced though an overhead due to the increased amount of messages used to maintain synchronization which has an impact on the energy consumption and the latency of the network. A novel TSCH schedule called Instant \cite{10.5555/3324320.3324325}, is utilized in indoor localization and data collection services in application scenarios designed for remote healthcare in residential environments. Instant includes both stationary and mobile nodes and the new feature offered in this schedule is that the mobile nodes are able to know information about reservations of subsequent unicast cell block of the stationary nodes in advance through a simple probe–ACK transaction. Therefore, if they want to transmit a large amount of data, they select a stationary node based on this information to select an available node and avoid delay and energy overhead. In a previous work we investigated how popular TSCH schedules perform heterogeneous mobility patterns \cite{iot2040033}. More specifically, we examined through simulations how MSF \cite{rfc4180}, Orchestra\cite{Duquennoy2015} and Alice\cite{Kim2019} schedules perform on mobility patterns representing an autonomous agricultural vehicles application scenario and a smart warehouse where a synergy between machinery, autonomous robots/vehicles and workers communicate. We illustrate that neither of the schedules is providing a reliable performance because a number of nodes are suffering from coverage issues or the schedule cannot handle the mobility and most of the times is outdated. 

An extended survey on a significant amount of mobility models and their impact on RPL protocol is presented in \cite{9187775}. The authors provide a comprehensive taxonomy, a classification of the mobility models and they conduct a comparison based on their main specification. Moreover, they evaluate RPL used in several mobility models using Cooja simulator to quantify metrics such as power consumption, reliability, latency and control overhead. An interesting observation is that the RPL's trickle timer, which is used to control traffic overhead by dividing the time into intervals, has an impact on the performance. Barceló et al. propose an extension of the 6TiSCH routing in \cite{6966074}, considering scenarios where static and mobile nodes co-exist. The routing between static nodes is carried out by the regular 6TiSCH but the routing among mobile nodes and static nodes is performed by a novel approach which is utilizing the end to end reliability estimations with a blacklisting function based on the position of the node. They evaluate their approach through simulations and the results show that the reliability between the mobile and static nodes is increased even when there are occurring high positioning errors compared to previous routing approaches. Another approach to improve mobile scenarios which use RPL is presented in \cite{9595872}. The authors follow a Software Defined Network (SDN) technique where a central controller has a holistic view of the network, can predict the handovers and update the routing tables. In addition, the central controller is not limited in terms of resources like the regular IoT nodes. Thus, it can execute computational intensive algorithms such as particle and Kalman filter. The authors evaluate their approach through Cooja simulations and the results suggest that the proposed approach can achieve higher reliability compared with the baseline RPL but also with the current state of the art mRPL.

\section{Active Connectivity}
\label{sec:active_connectivity}
The AC approach entails several challenges depending on the application requirements. This section is establishing the foundations of enabling AC in TSCH networks with mobile nodes. The main principle is that if a link between two mobile nodes is weak and is about to die then the nodes have the ability to regulate physical characteristics like their speed, their topology or their route to keep this link alive. In this paper we focus only on the speed due to limited space but still this is very general principal that generates a set of technical concerns that have to be addressed. 

The first consideration to be taken is to decide which of the two nodes is going to regulate its speed. The routing in TSCH networks is organized with a tree topology most of the times and we decided that it will be convenient that the child node will be the one which will regulate its speed because it will probably affect less nodes in the network. In that sense if the child node has its own children, then these nodes have to regulate their speed as well in the same rate to remain within the connectivity range. Another consideration was to select the metric which would indicate that a link is about to die. The IoT devices are using low cost radios which are prone to noise and interference \cite{10.1109/IPSN.2008.66,8115772}. There are metrics like Radio Signal Strength Indicator (RSSI) which is the captured power level after a packet reception. The RSSI measurement unit is decibels per milliwatt (dBm). A similar metric is the Link Quality Indicator (LQI) which describes the quality of the received packet but it is less depended by the environment activity compared with RSSI. Both of them have significant spatial or temporal dependencies especially during a mobile scenario. This makes it challenging to find a balance between a representing quantity of samples and be coherent with the limited amount of resources offered by IoT devices. To this end we decided to determine a weak link based on the Exponentially Weighted Moving Average (EWMA) of RSSI values, which is described in  Equation $\ref{eq:ewma}$.
\begin{equation}
    EWMA_{RSSI_{i}} = \alpha * RSSI_{i} + (1-\alpha) * EWMA_{i-1} 
    \label{eq:ewma}
\end{equation} 

\begin{figure}[tp]
    \centering
    \includegraphics[scale=.55]{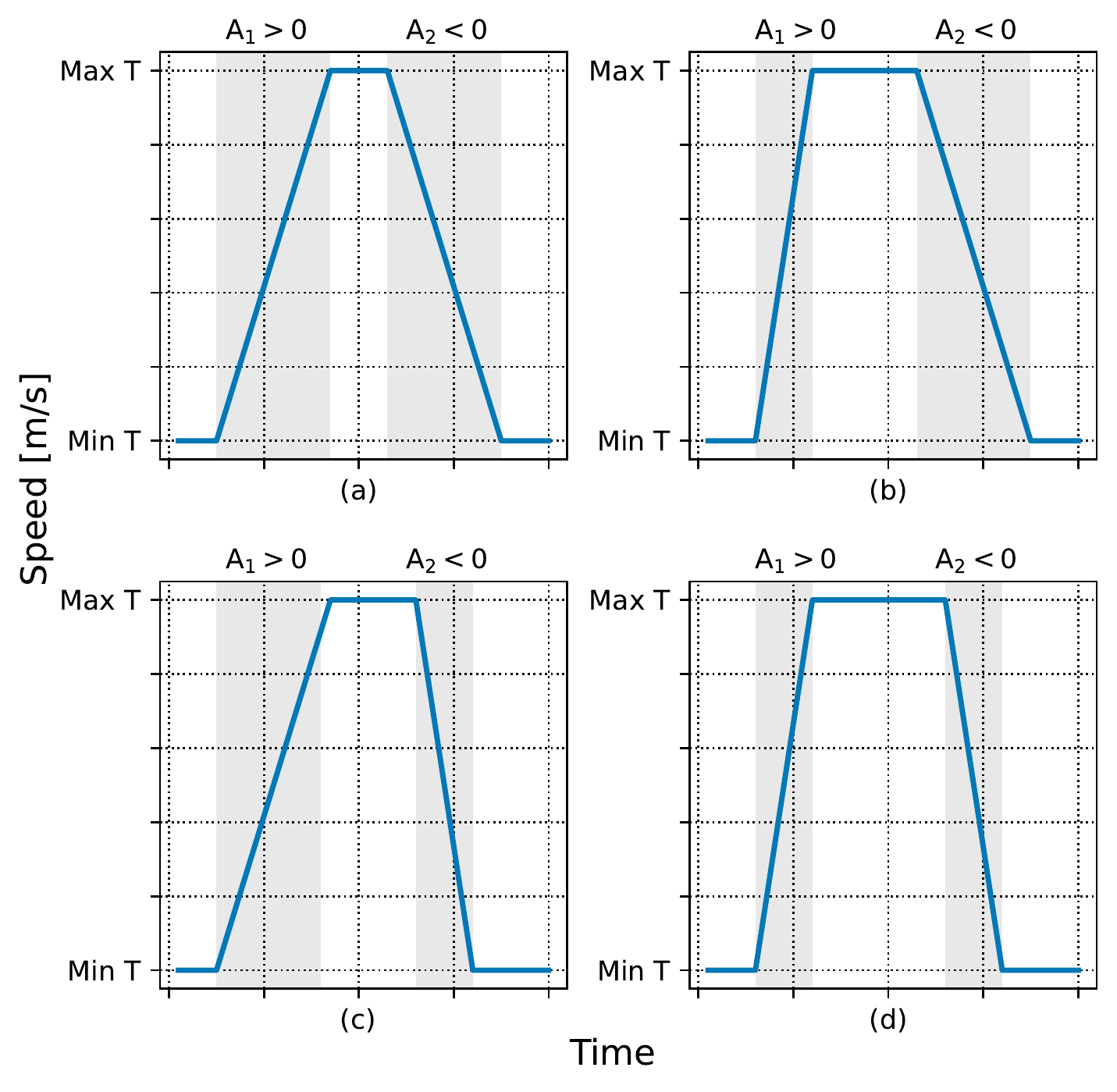}
    \caption{Representation of potential speed profiles. Example of a moderate acceleration and deceleration adaptation (a), within the gray intervals the acceleration $A_{1}$ and deceleration $A_{2}$ represent the speed change, which is activated based on the Minimum and Maximum RSSI Thresholds. In (b) it is denoted the steep acceleration moderate deceleration, in (c) the moderate acceleration steep deceleration and in (d) the steep acceleration steep deceleration technique.}
    \label{fig:active_connectivity}
\end{figure}

EWMA is not consuming a lot of resources and at the same time it is able to represent a series of RSSI values. The selection of the alpha parameter, which is defining the weight of the current RSSI value and the weight of the previous RSSI values, is examined in the evaluation part in Section \ref{sec:evaluation} along with different RSSI thresholds used to notify the prospect of a weak link. The rate that the speed is going to be regulated was also one of the considerations we had to take. Apparently the acceleration or the deceleration can affect the performance of the AC operation. Especially when a node is at the very limits of the connectivity range it is important to regulate its speed faster. We illustrate some potential speed profiles we in Figure $\ref{fig:active_connectivity}$. Namely, a moderate approach both for acceleration and deceleration illustrated in Figure $\ref{fig:active_connectivity}$a, an approach which includes a steep acceleration and a moderate deceleration in Figure $\ref{fig:active_connectivity}$b, another one where the acceleration is moderate and the deceleration steep in Figure $\ref{fig:active_connectivity}$c and one where both acceleration and deceleration are steep in Figure $\ref{fig:active_connectivity}$d. In this paper we focus only at the speed profile illustrated in $\ref{fig:active_connectivity}$a for practical reasons.

The metrics to evaluate the performance of AC have to be considered as well. Networking metrics like throughput, packet delivery ratio are able to represent the network performance of the TSCH network but we also use the downtime which we define as the time a node spends not being connected to the TSCH network after the first time it gets connected. Moreover, the AC will have an impact on the distance a node will cover. Therefore, we consider the moving distance as a metric towards the application layer, because if a node covers a shorter distance due to AC purposes, it might increase the operating cost since the node would require more time to operate. For instance, if the node is an autonomous vehicle, it would require more fuel in that case.

\section{Evaluation}
\label{sec:evaluation}
\begin{figure}[t]
    \centering
    \includegraphics[scale=.6]{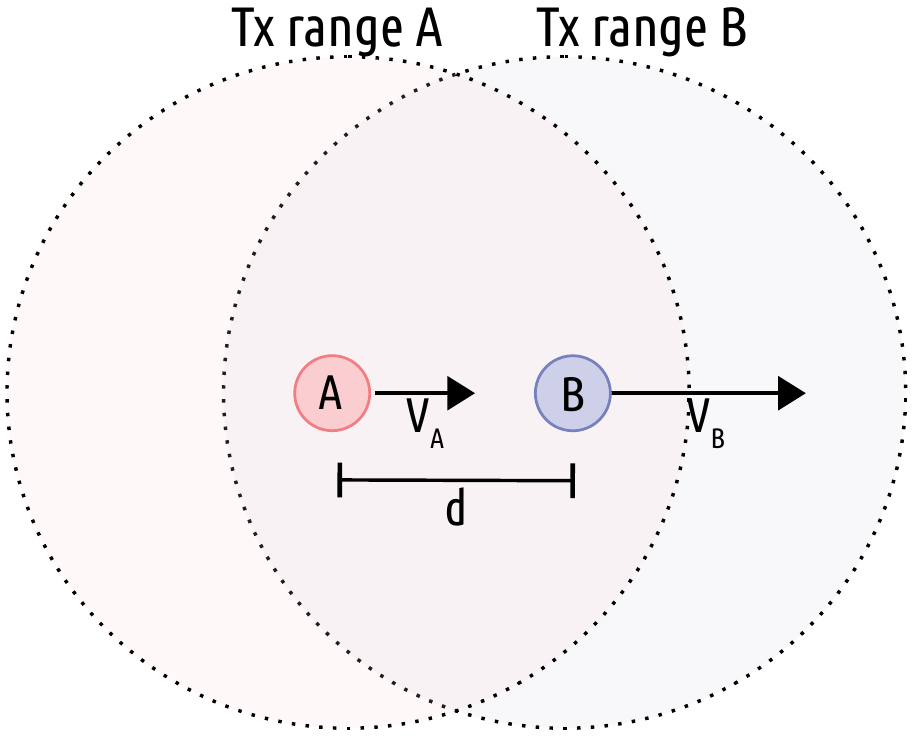}
    \caption{The rudimentary mobility pattern Node A and Node B follow to model the design principles of the Active Connectivity mechanism. $3V_{A}=V_{B}$  }
    \label{fig:AC_mobility}
\end{figure}

This section describes all the technical details during the evaluation part, the obtained results and a discussion upon them.
In order to evaluate the AC we used the Cooja simulator \cite{cooja} and we defined a simplistic topology with two nodes depicted in Figure $\ref{fig:AC_mobility}$. The mobility model we use is also very simple, node A in Figure $\ref{fig:AC_mobility}$ is the root node and it is moving to the same direction with node B with a distance of $130$ m. They are moving in a straight line form until they reach the end of the grid and then they return back. They repeat this pattern for $30$ minutes of simulation time. Node A is moving with $1$ m/s and node B with $3$ m/s. The maximum speed a node can go is up to $3$ m/s and the minimum $0$ m/s. We define these regulations for safety reasons (a mobile node can be a vehicle, a robot, a UAV and cannot move too fast) but also to consider the covered distance as a meaningful metric to evaluate the AC impact on the application layer.  All the simulations were repeated with $20$ different random seeds to avoid statistical bias. The configuration parameters of the simulation are listed in Table \ref{tab:sim_parameters}.

We evaluate two versions of AC mechanism. In the first one we assume that a node knows the position of the rest of the nodes in the network, thus when AC notifies it about a weak link it knows if it has to accelerate or decelerate. We do this assumption due to the fact that several application scenarios such as autonomous agricultural vehicles most probably have GPS and can aggregate their position through the TSCH network but also in the spirit of abstracting complexities. We refer to this version as AC. In the other version a node does not know the positions of the other nodes and the mechanism does a random guess to decide if it has to accelerate of decelerate when it receives a notification about a weak link. This decision is evaluated during the operation and if the EWMA level drops down to threshold after this decision has been taken, the mechanism changes its decision and does the opposite action. We refer to this mechanism as AC Random (ACR). In the ACR case the topology depicted in Figure $\ref{fig:AC_mobility}$ could change if node B decrease a lot its speed and node A overtake it and it depends up to the configurations parameters that node B increase its speed in time to continue being in range with node A. We also run a version with the AC mechanism deactivated to have it as a benchmark comparison.

\begin{figure*}[tp]
    \centering
    \includegraphics[scale=.5]{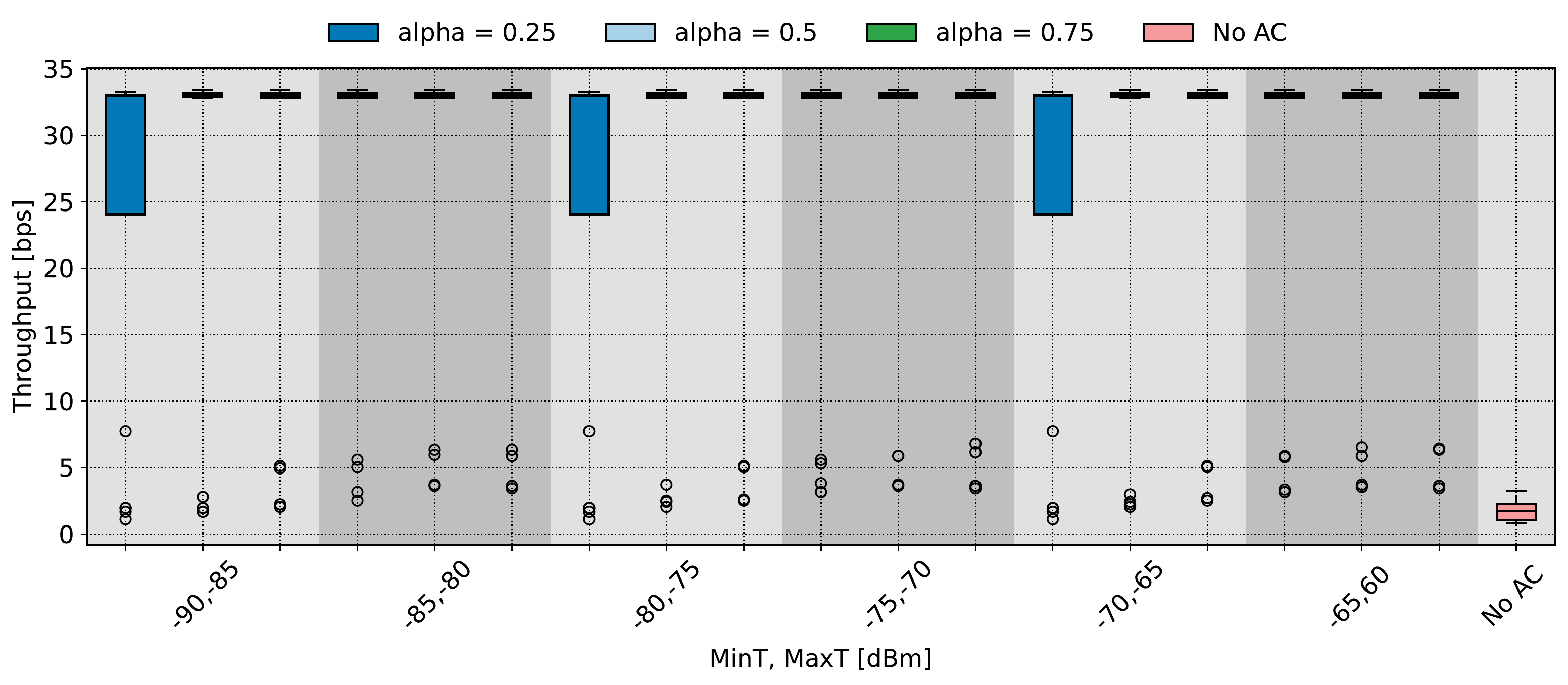}
    \caption{Throughput using AC for different alpha values and RSSI thresholds. The colors representing the alpha values are not visible because the throughput performance is homogeneous close to its maximum value for most of the cases.}
    \label{fig:throughput_ac}
\end{figure*}

\begin{figure*}[tp]
    \centering
    \includegraphics[scale=.5]{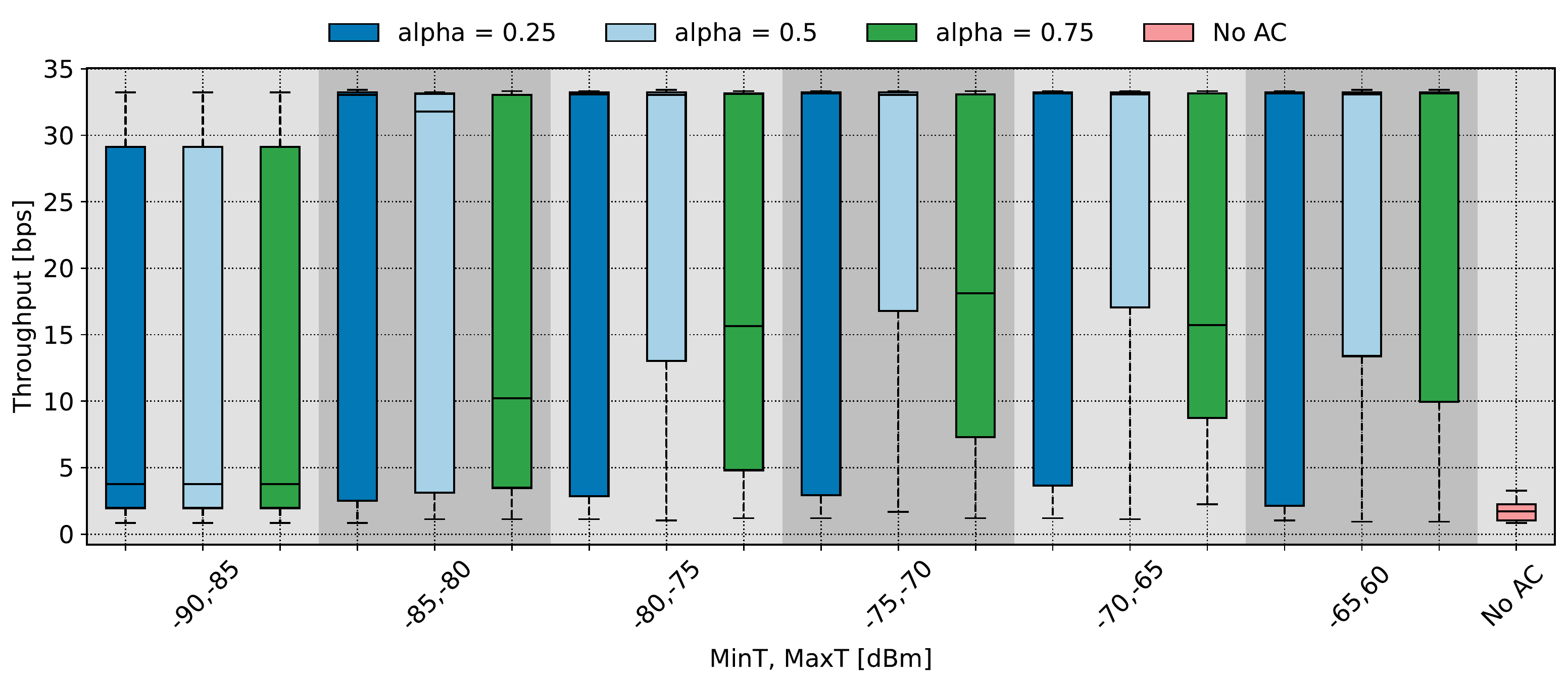}
    \caption{Throughput using ACR for different alpha values and RSSI thresholds.}
    \label{fig:throughput_acnogps}
\end{figure*}

\begin{table}[]
\centering
\caption{Configuration parameters used during the simulation}
\begin{tabularx}{0.48\textwidth}{ |>{\centering\arraybackslash}X | >{\centering\arraybackslash}X| }
\hline
Parameter & Value \\ \hline
Grid size & 2000 x 2000 m  \\
Transmission range & 450 m \\
Radio model & Logistic Loss  \\
Mobility model & Mobile Ad Hoc Network\\ 
Mobility speed & 0 to 3 m/s  \\
Traffic rate & 1 packet/5 sec \\
Number of nodes & 2 \\
Slotframe length & 3\\
Slot duration & 10 ms \\
TSCH schedule & Minimal Scheduling Function \\

Transceiver & CC2420 \\
Frequency spectrum & 2.4 GHz \\
Simulation duration & 30 min \\ \hline
\end{tabularx}
\vspace{0.25cm }
\label{tab:sim_parameters}
\end{table}

Figures $\ref{fig:throughput_ac}$ and $\ref{fig:throughput_acnogps}$ illustrate the first set of results which is the achieved throughput. Thus, Figure $\ref{fig:throughput_ac}$ presents the throughput for three different alpha values used in EWMA and six different RSSI threshold values which will activate the AC mechanism. Notice that the RSSI threshold values are mentioned in couples as the lower will activate the AC mechanism when the link is weak and the higher when the link is in a better status and regulate its speed towards to its initial speed. All the following Figures present the results on this manner.

AC was able to achieve approximately $33$ bps for most of the cases which is the maximum it can be achieved in this simulation. However when the alpha value was $0.25$ there are few cases when the throughput was lower close to $25$ bps. We speculate that this due to the manner EWMA functions, namely with a lower alpha value the weight of the last RSSI value is lower in the EWMA equation. Hence, AC requires more time to respond to weak link case.

Figure $\ref{fig:throughput_acnogps}$ shows the throughput for the ACR case. The throughput is lower for some cases, for instance when the minimum and maximum thresholds were $-90$ and $-85$ dBm. An explanation might be that the high values in the thresholds notify ACR about a weak link too late and if the random decision is mistaken there is no time to change the decision before the node goes out of range. When the alpha is $0.75$, ACR performance drops. The explanation in this case is that alpha value increases a lot the weight of the last RSSI value making ACR too aggressive which in combination with the random guesses increases the mistaken decisions. In general even though ACR performs worse than the AC version which is expected but if we compare the ACR with the version without AC features there is a significant improvement.

\begin{figure}[tp]
    \centering
    \includegraphics[scale=.51]{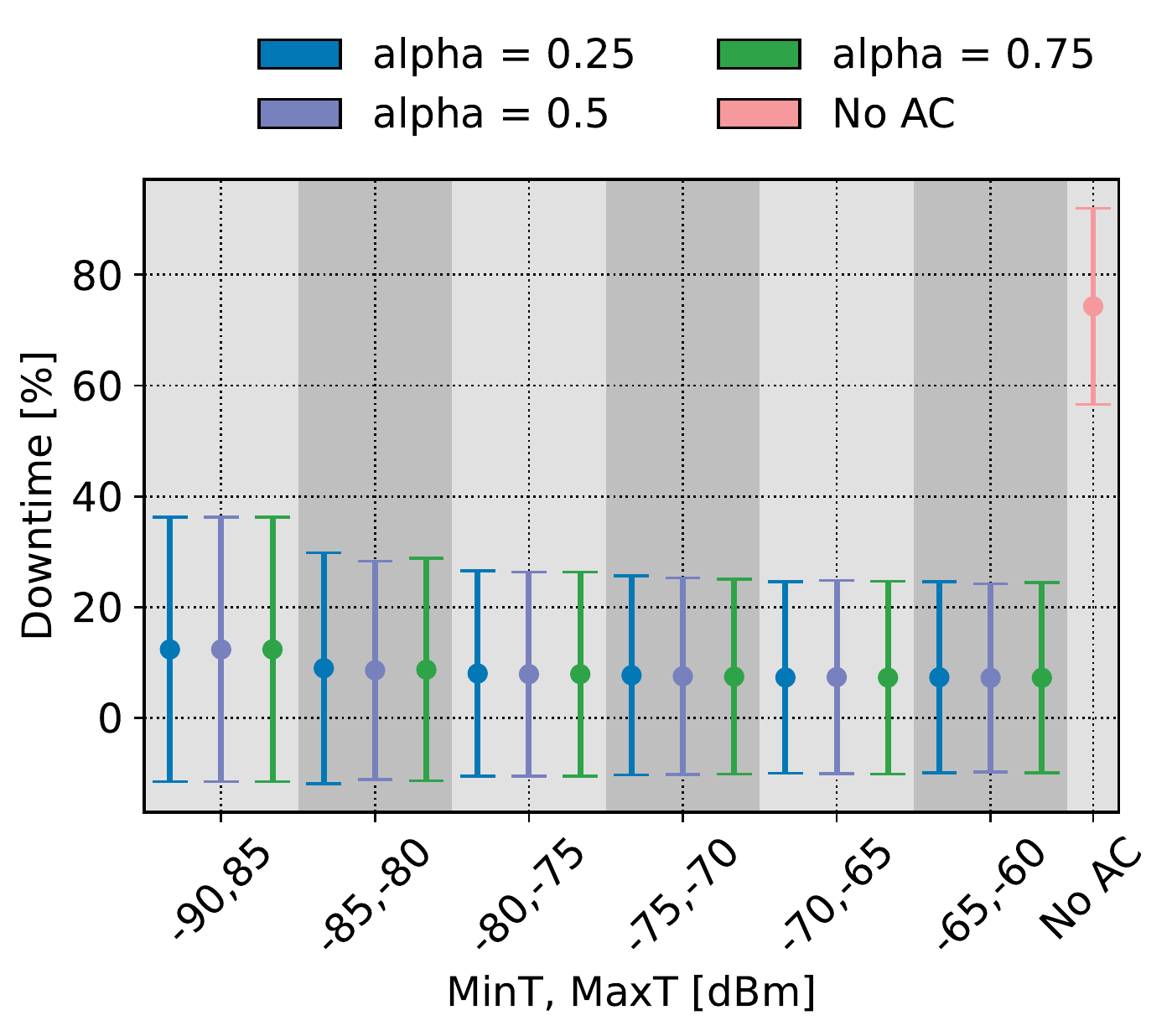}
    \caption{Downtime using AC for different alpha values and RSSI thresholds.}
    \label{fig:downtime_ac}
\end{figure}

\begin{figure}[tp]
    \centering
    \includegraphics[scale=.51]{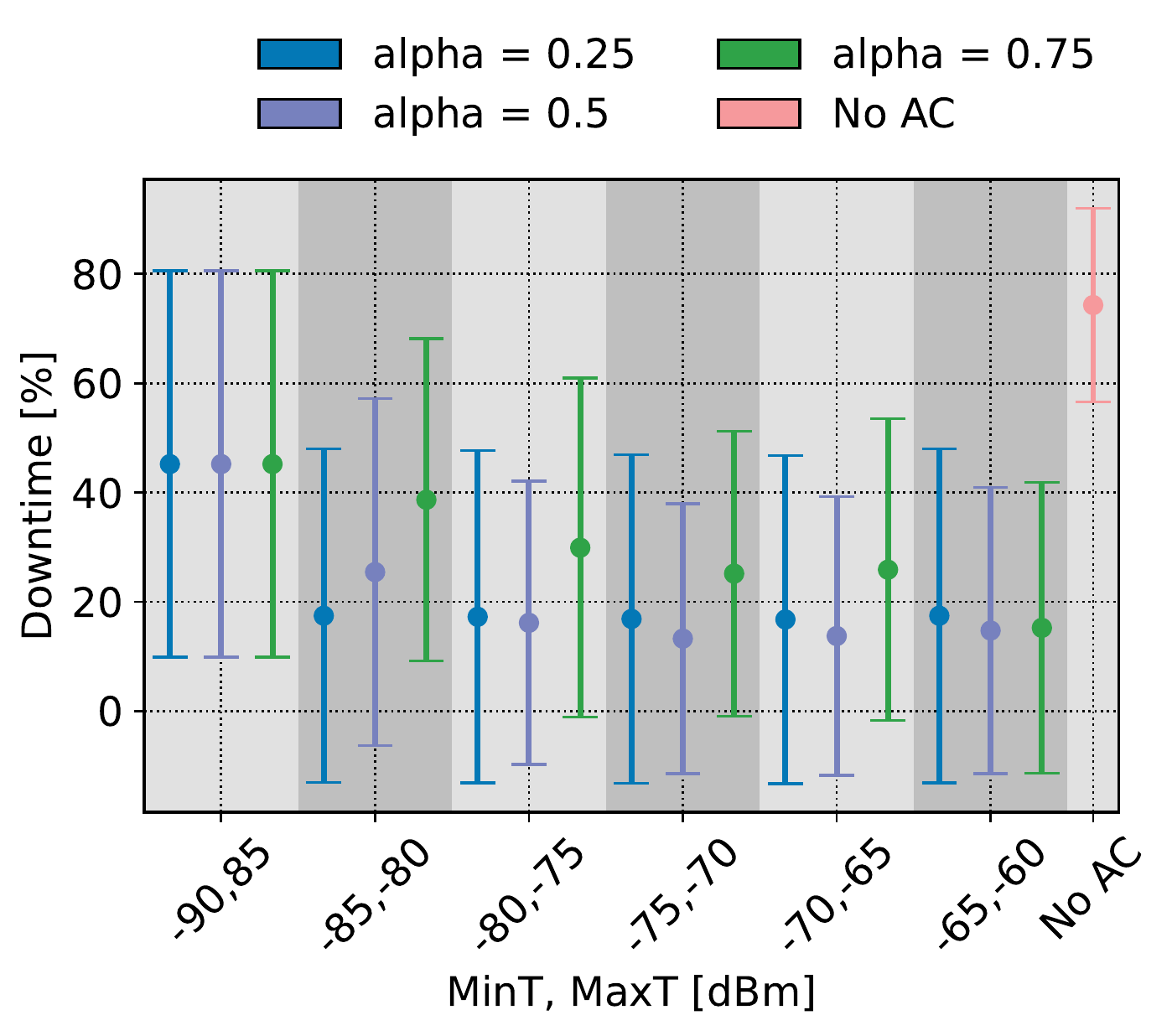}
    \caption{Downtime using ACR for different alpha values and RSSI thresholds.}
    \label{fig:downtime_acnogps}
\end{figure}

In Figures \ref{fig:downtime_ac} and $\ref{fig:downtime_acnogps}$ we see the downtime of AC and ACR. The bullet points here stand for the mean value and the error bar for the standard deviation. The mean downtime for AC was approximately $10 \%$ and the standard deviation approximately $20$. Apparently, the trends are the same with the throughput since they are obtained from the same simulations. For the ACR case we see higher percentages of downtime, the alpha that achieves the lowest downtime, considering all the threshold values is $0.25$. If we compare the version without AC features with the worst performance of ACR when thresholds were $-90$ and $-85$ dBm, the downtime is reduced by $30\%$ approximately.

\begin{figure}[tp]
    \centering
    \includegraphics[scale=.5]{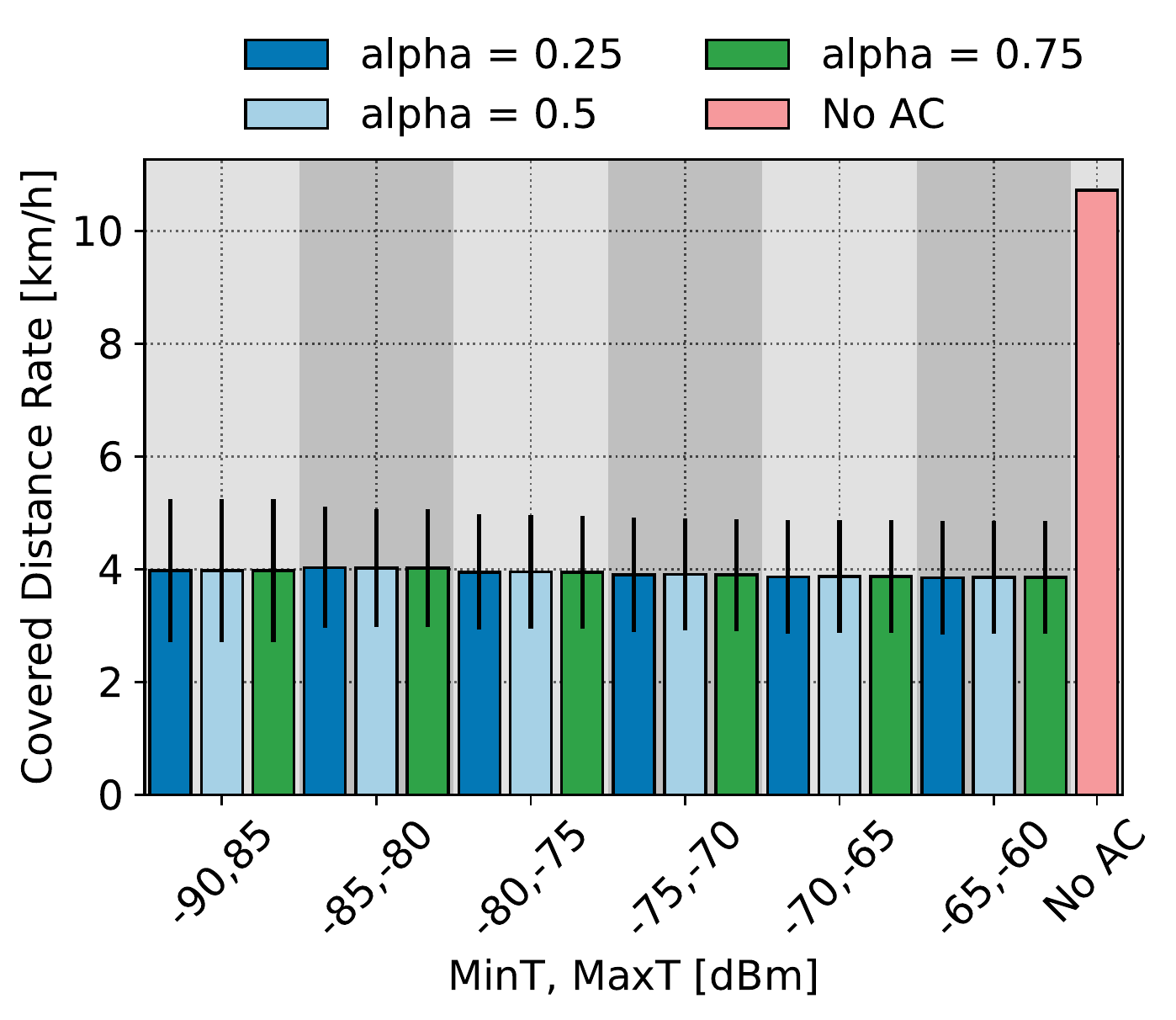}
    \caption{Distance using AC for different alpha values and RSSI thresholds.}
    \label{fig:distance_ac}
\end{figure}

\begin{figure}[tp]
    \centering
    \includegraphics[scale=.5]{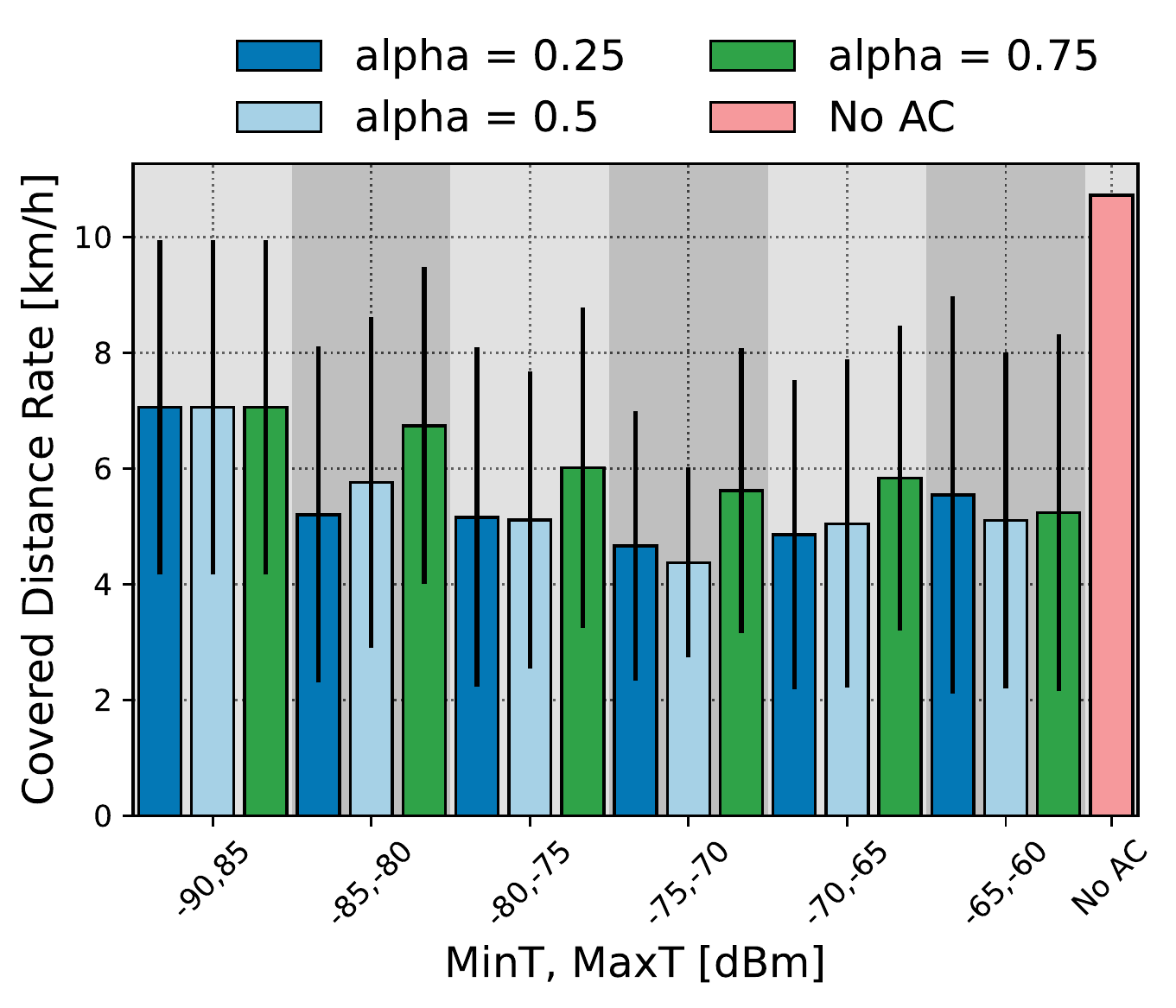}
    \caption{Distance using ACR for different alpha values and RSSI thresholds.}
    \label{fig:distance_acnogps}
\end{figure}

The distance covered by node B is represented in Figures $\ref{fig:distance_ac}$ and $\ref{fig:distance_acnogps}$ for the case of AC and ACR respectively. The bars represent the mean distance and the errorbars the standard deviation. We decided to use this performance metric as well since it can have on impact on the overall cost. The distance node B covers without AC features is around $11$ km/h and when AC is activated in Figure $\ref{fig:distance_ac}$ the distance drops to $4$ km/h since it has to reduce its speed to be in range with node A. This is an overhead that some applications might not be able to afford. Respectively, some applications might be able to afford a lower throughput in exchange with more covered distance. This is achieved by ACR in Figure $\ref{fig:distance_acnogps}$ where the mean covered distance is $7$ km/h when the threshold values were $-95, -85$. This happened because node B spent more time disconnected from the TSCH network of course as it is indicated from Figure $\ref{fig:downtime_acnogps}$. The tradeoff described from the results is that the throughput is disproportionate with the downtime and the covered distance.

\section{Conclusion}
\label{sec:Conclusion}

We introduced AC approach for TSCH networks including mobile nodes. AC can regulate the speed of the nodes to increase the reliability by keeping the nodes associated within the TSCH network by notifying them when they are going out of coverage. Utilizing Cooja simulator and generating a simplistic scenario we manage to quantify the performance of two versions of AC. One where a node knows the position of the other nodes and therefore knows if it should accelerate or decelerate its speed to within range and another one where the node does not know and have to guess using a random function. The results show that the reliability can be increased but there is an impact on the distance and consequently on the application performance. We describe this tradeoff and discuss that depending the application requirements it can be crucial or not. In Future work we plan to evaluate AC with more complex topologies and real experiments including smart car kits and real IoT nodes. Moreover, the case of ACR would be investigated if it can be implemented with Reinforcement Learning methods \cite{kober2013reinforcement} to replace the random guessing.  

\clearpage
\bibliographystyle{ieeetr}
\bibliography{references}

\end{document}